\documentclass[12pt]{iopart}
\usepackage{graphicx}
\usepackage{hyperref}

\newcommand{\amuHLO }{$a_\mu^{\rm{HVP,\,LO}}$}

\begin{document}

\title{Status of the MUonE experiment}

\author{G Abbiendi\footnote{for the MUonE Collaboration} }

\address{INFN - Sezione di Bologna, Viale Carlo Berti-Pichat 6/2, 40127 Bologna, Italy}

\ead{giovanni.abbiendi@bo.infn.it}

\date{\today}

\begin{abstract}
The MUonE experiment has been proposed to measure the differential
cross section of $\mu e$ elastic scattering, by colliding the 160 GeV
muons of the CERN M2 beam with atomic electrons of thin target plates.
From a very precise measurement of the shape one
can achieve a competitive determination of the leading hadronic contribution to the muon magnetic
moment, independent from the other existing ones.
In preparation for the Test Run with a reduced setup the detector
geometry has been optimised.
Expected yields for a first physics run with limited statistics are
discussed, together with prospects for the assessment of the main systematic
uncertainties.
\end{abstract}


\maketitle

\section{Introduction}
The muon magnetic moment anomaly $a_\mu = (g_\mu -2)/2$  is one of the most precisely
measured quantities as well as one of the most precisely calculable in
the Standard Model (SM), therefore it constitutes a stringent test of the theory.
For the last twenty years the BNL measurement~\cite{Bennett:2006fi}
has been pointing to a significant discrepancy from the theory
prediction.
Recently the first result from the FNAL g-2 experiment~\cite{FNAL-g-2} has confirmed
the previous measurement,
and their combination brings to 4.2~$\sigma$ the
deviation from the currently accepted SM prediction \cite{Aoyama:2020ynm}.
The theory prediction is a formidable achievement of the Standard
Model, including terms up to five loops of perturbation theory for the
dominant QED part, with significant contributions coming also from
weak interactions and QCD. An extensive review of the state-of-the-art
calculations is given in \cite{Aoyama:2020ynm}.

Many hypotheses have been put forward to explain this deviation as a quantum
effect of new exotic particles, however it is also possible that it
result from a systematic error in the calculation.
The dominant uncertainty of the prediction comes from the leading order contribution from hadronic
vacuum polarisation \amuHLO, which is not calculable in
perturbation theory. This is usually determined by a data-driven
dispersive approach, using the low-energy measurements of hadronic
production in $e^+e^-$ annihilation \cite{Davier:2019can, Keshavarzi:2019abf}.
In contrast, a recent ab initio calculation of \amuHLO ~based
on Lattice QCD \cite{BMW20} reduces the discrepancy with the
measurement, and is in tension with the data-driven estimates.

In the next years the FNAL experiment is expected to increase the
precision by about a factor of 4, while another forthcoming experiment
at J-PARC \cite{Abe:2019thb} should reach a similar precision.
Therefore it is of paramount importance to improve the theory
calculation and clarify the comparison of the available estimates.
A novel approach has been proposed in \cite{Calame:2015fva}, to determine the leading hadronic
contribution \amuHLO~ from a measurement of the effective
electromagnetic coupling in the space-like region,
where the vacuum polarisation is a smooth function.
It is based on the equation \cite{deRafael}:
\begin{equation}
 \label{amu_xalpha}
        a_\mu^{\rm{HVP,\,LO}} = 
         \frac{\alpha}{\pi} \int_0^1 dx \, (1-x) \,  \Delta \alpha_{\rm had} \! \left[ t(x) \right],
\end{equation}
where $\Delta\alpha_{\rm had}(t)$ is the hadronic contribution to the
running of the QED coupling, evaluated at the space-like (negative) squared four-momentum transfer:
\begin{equation}
        t(x)=-\frac{x^2m_\mu^2}{1-x} < 0.
\label{t}
\end{equation}
The running QED coupling is expressed as:
\begin{equation}\label{eq:alphaq2}
        \alpha(t) = \frac{\alpha(0)}{1-\Delta \alpha(t)},
\end{equation}
where $\alpha(0)=\alpha$ is the fine-structure constant, and
\begin{equation}\label{eq:Deltalpha}
        \Delta \alpha(t) = \Delta\alpha_{\rm lep}(t) + \Delta\alpha_{\rm had}(t).
\end{equation}
The hadronic contribution $\Delta\alpha_{\rm had}(t)$ can be extracted
by subtracting from $\Delta \alpha(t)$ the purely leptonic part
$\Delta \alpha_{\rm lep} (t)$, which can be calculated to very high precision in QED.

Very few direct measurements of the running of $\alpha$
in the space-like region exist to date. The most precise one was obtained by the OPAL
experiment \cite{opal}, from small-angle Bhabha scattering at LEP, and reached
the sensitivity for the observation of the hadronic contribution.

Based on the method of \cite{Calame:2015fva}, the MUonE experiment \cite{muone}
has been proposed, to measure the hadronic running of $\alpha(t)$ 
from $\mu e$ elastic scattering at low energy.
This method could reach a competitive precision below $0.5\%$ on the
\amuHLO, provided the systematic errors are kept under control.
The MUonE project has been
submitted to the CERN SPS Committee with the Letter-of-Intent
\cite{LoI} in 2019.
A Test Run has been approved with a partial apparatus as a validation of the
detector design and of the overall concept.

In this paper the current status of the project is summarised, and some
recent developments are described.
Section~\ref{sec:detector} describes the proposed MUonE setup
with a recent geometry optimisation. Section~\ref{sec:analysis}
reviews the analysis technique in some detail, and
section~\ref{sec:TestRun} discusses the achievable yields in
the forthcoming Test Run, highlighting the sensitivity to the main
systematic effects which have to be dealt
with. Section~\ref{sec:theory} lists the main theoretical
advancements which constitute the other important face of the
challenge. 
Finally the conclusions are given in section~\ref{sec:conclusions}.

\section{MUonE experimental apparatus}\label{sec:detector}
\begin{figure*}[!htbp]
\begin{center}
\includegraphics[scale=0.35]{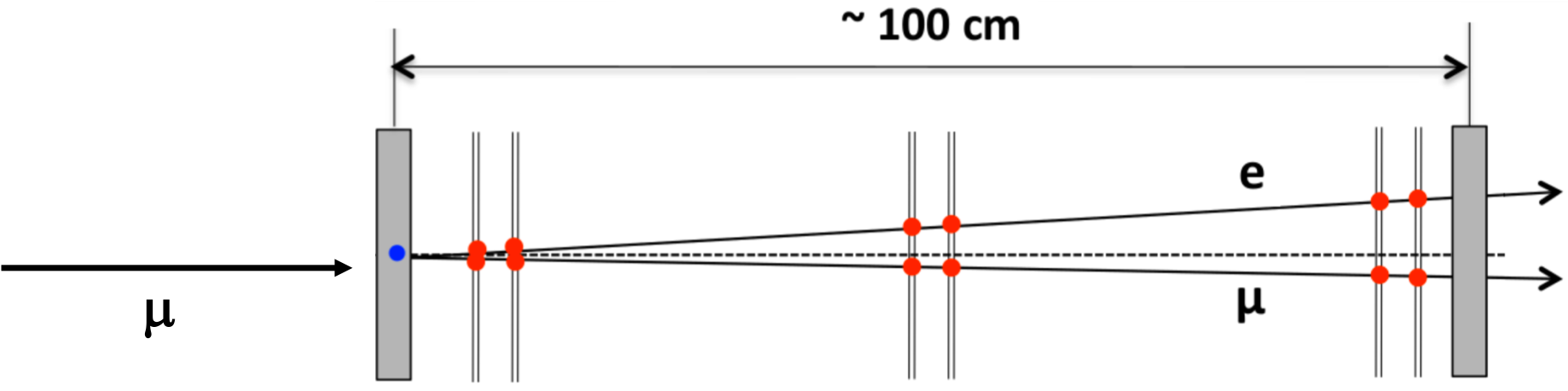}
\caption{Scheme of one tracking station.}
\label{station}
\end{center}
\end{figure*}

\begin{figure*}[!htbp]
\begin{center}
\includegraphics[scale=0.8]{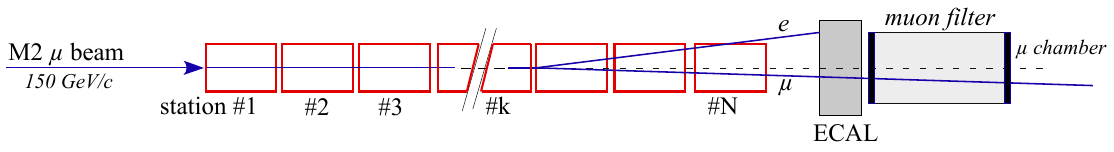}
\caption{Layout of the MUonE experimental apparatus (not to scale).}
\label{det-layout}
\end{center}
\end{figure*}

The idea of the MUonE experiment has been put forward in \cite{muone}.
The hadronic running of the QED coupling needed in the master equation (\ref{amu_xalpha})
can be determined by a very precise measurement of the shape of the
differential cross section of $\mu e$ elastic scattering, using the
CERN M2 muon beam ($E_\mu \sim 150$-$160$~GeV) off atomic electrons of a
light target. This process has several attractive features:
\begin{itemize}
\item simple kinematics;
\item pure $t$-channel;
\item useful centre-of-mass energy to probe the dominant region for
  the muon $g-2$;
\item easy selection based on the correlation of the electron and muon scattering angles.
\end{itemize}
The proposed detector has been described in \cite{LoI}.
The scattering angles of muons and electrons are measured by
tracking stations as the one represented in Figure~\ref{station}.
An elastic scattering event produced in the thin target plate in front of the
station, with thickness of $1.5$~cm, is identified by measuring the two
outgoing tracks in three pairs of planes made of silicon microstrips
with orthogonal strips, over a length of one meter. The tracking
precision is crucial, so in addition to a good intrinsic
resolution it is necessary to limit the multiple Coulomb scattering. 
For this purpose the target has to be made of a low-Z
material as beryllium or carbon and has to be thin. To reach the
necessary interaction rate while preserving these conditions a modular
layout is proposed, forming an array of identical
stations crossed by the muon beam, as shown in
Figure~\ref{det-layout}.
The full apparatus consists of 40 such stations, followed by an
electromagnetic calorimeter (ECAL) and a muon detector at the end, to help
the identification and the selection.
Beam muons are almost unaffected by the upstream detector material, except for a small energy loss, so
every station behaves as an independent detector. The events occurring
at a given station have the incoming muon direction measured by the
preceding station.
This configuration could reach
the target statistical sensitivity in three years of run at the M2
beam, collecting an integrated luminosity of $1.5 \times 10^{7}$~nb$^{-1}$ \cite{LoI}.

The basic tracking unit has been chosen to be the 2S module
developed for the upgrade of the CMS outer tracker
\cite{TRACKER_UPG_TDR}. It consists of two close-by planes of silicon
microstrips, separated by 1.8~mm, with strips along the same direction
in the two sensors and reading the same coordinate.
A pair of matching hits in the two sensors gives a so-called {\em
  stub}, or track element, providing track triggering capability at $40$~MHz, with inherent suppression of
background from single-layer hits or large-angle tracks.
It has a large active area of about $10\times10 \mathrm{~cm}^2$,
allowing to completely cover the relevant MUonE angular acceptance
with a single module, thus assuring the best uniformity.
With respect to the LHC operation the main
difference for MUonE will be the asynchronous nature of the signals
from the $\mu e$ scattering events. This will be managed by a
specific configuration of the front-ends, and will be studied in
detail during the Test Run.

\subsection{Detector optimisation}
\begin{figure*}[!htbp]
\begin{center}
  \includegraphics[scale=0.3]{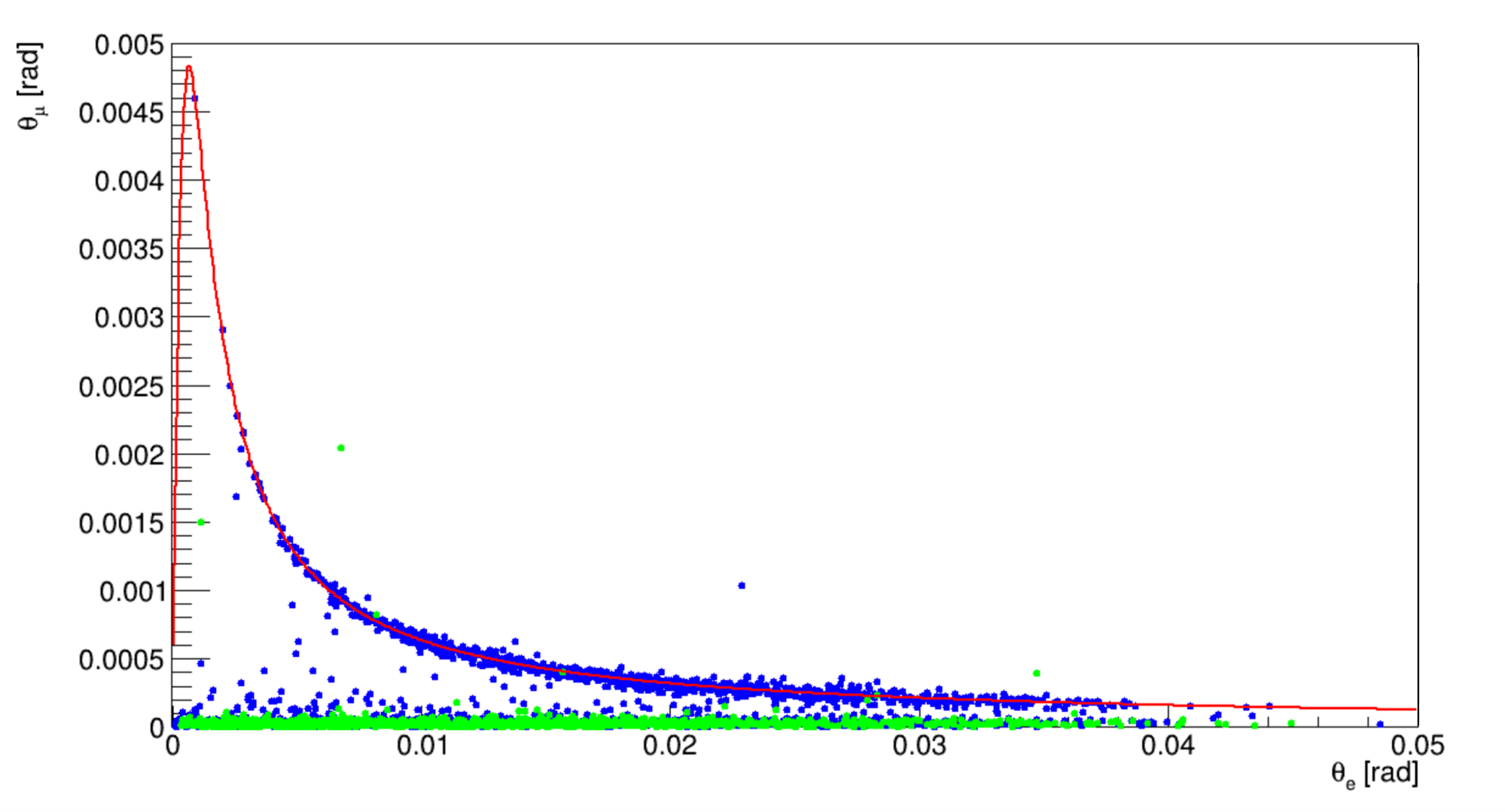}
  \includegraphics[scale=0.3]{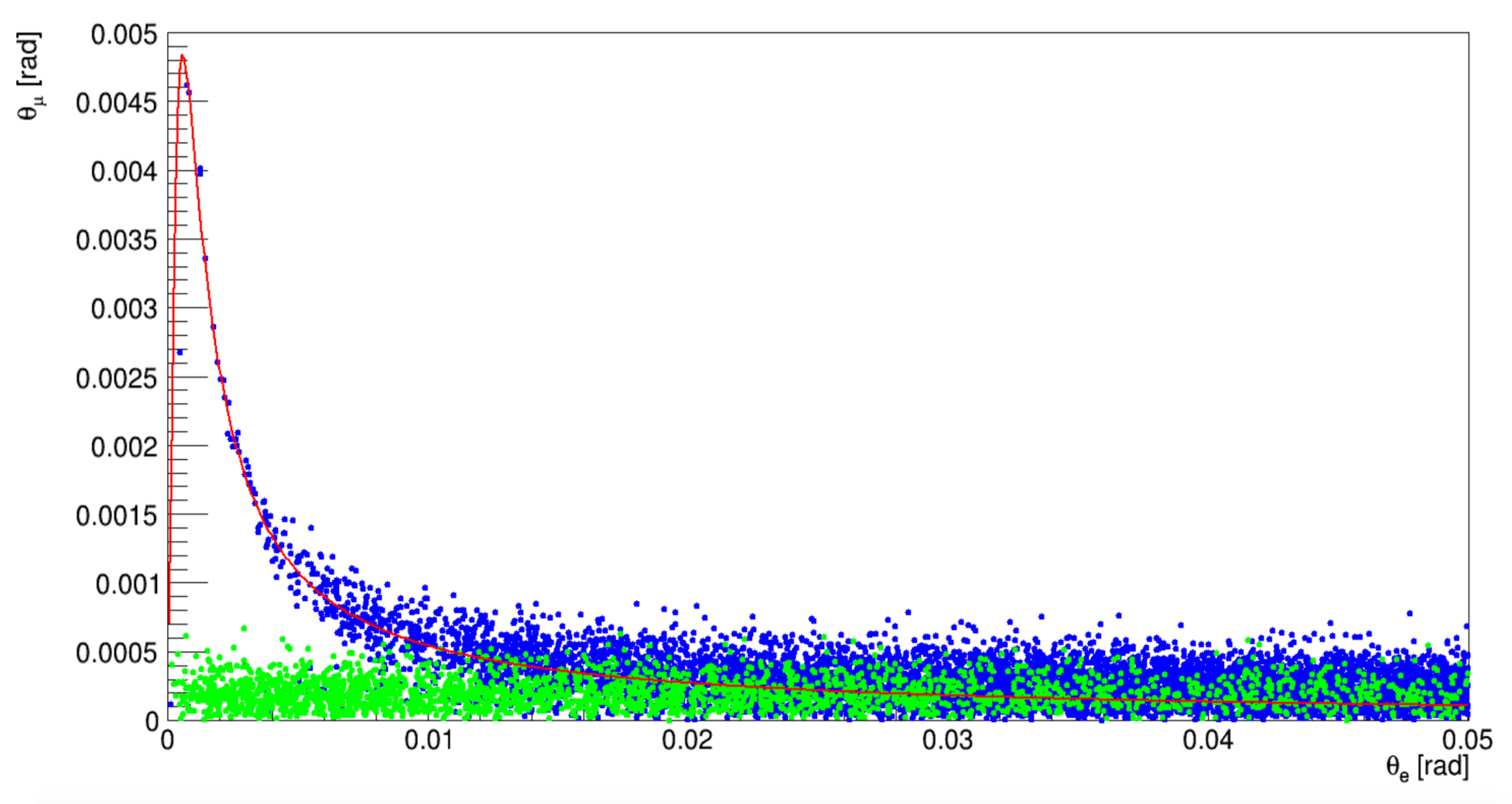}
\caption{Simulated angular distributions of two-track events:
  (\emph{Top}) for the 2017 beam test with the UA9 detector
  \cite{TB2017};
  (\emph{Bottom}) for the 2018 beam test \cite{TB2018}.
  The blue points represent $\mu e$ events, the green ones background
  events with production of an $e^+e^-$ pair in the material. The red
  curve represents the ideal elastic scattering. Plots from \cite{MatteoBonanomiThesis}.
}
\label{TB17-18_sim}
\end{center}
\end{figure*}

The hit intrinsic resolution of the tracking detector is particularly
important. This is clearly demonstrated by Monte Carlo simulations
based on GEANT4 \cite{GEANT4, GEANT4-recent-develop}. Figure~\ref{TB17-18_sim} shows the
distributions obtained for the scattering angles of the muon and the
electron, reconstructed from the simulated hit patterns, corresponding
to two different tracking setups. In both cases the reconstructed
events include both the signal (elastic scattering events) and
background events from $e^+e^-$ pair production, which can mimic
elastic events when one electron goes undetected. The top plot
corresponds to the performance obtained in a beam test with the UA9
detector \cite{TB2017}, which achieved a position resolution of
$7\mu$m on individual hits.
The bottom plot corresponds to the performance obtained in another beam test
\cite{det2018,TB2018}, which featured a worse resolution of $\sim 35$-$40 \mu$m.
A good intrinsic resolution is crucial to allow for an effective
separation of the signal from the background by cutting on the muon
scattering angle at $\theta_\mu \geq 0.1-0.2$~mrad.

The CMS 2S module has a strip pitch of $90\mu$m and operates
with digital readout. Hence hits recorded from single-strip clusters will have a resolution
of about $90\mu$m$/\sqrt{12} \simeq 26 \mu$m on a single detection
plane, which does not seem optimal for MUonE.
Possible improvements were suggested by an independent study taking
the MUonE detector as a case study \cite{dorigo}.
Following these hints a detailed study was carried out by a full simulation describing
both the geometry and the digitisation of the 2S module. 

The ionisation charge produced by the passage of a charged particle
through the silicon layer is collected on the
strips, giving a digital signal when it exceeds a configurable
threshold, which is typically set around 6 times the RMS
electronic noise level.
A well-known method to improve
the resolution of silicon strip detectors consists in tilting the
planes around an axis parallel to the strips. This was recently
applied to the CHROMIE telescope \cite{CHROMIE}.
This simple geometry change increases the probability of
charge sharing between adjacent strips and can produce a significant
number of clusters with two strips above threshold, which results in a
better estimate of the crossing point of the track.
The optimal working point is expected to correspond to an average cluster
width of 1.5 strips, which happens when an equal amount
of one- and two-strip clusters is found.
An additional improvement can be obtained by an effective
staggering of the two sensor layers constituting the 2S module. 
In fact a micro-tilt of 25~mrad is equivalent to an half-strip
staggering of the two sensor layers. The two effects, charge sharing
between adjacent strips of the same layer and staggering of the two
sensor layers constituting the 2S module, can sum up to obtain a
larger improvement.
The search for the optimal working point was carried out by comparing several different
simulations obtained for different values of the signal threshold and
the tilt angle. 
The best result was obtained for a tilt angle of 233 mrad and signal threshold at 6 times
the RMS noise level, with a resolution of $8.0\mu$m.
For comparison, the resolution for the non-tilted
geometry, with the sensors orthogonal to the beam direction, was found
to be about $22 \mu$m. The study also tested the robustness of the
result under mechanical imperfections in the 2S module assembly, in
particular a possible misalignment of the two sensor layers along the
measurement coordinate, orthogonal to the strip direction, which would
be equivalent to an unwanted staggering of the two layers.
Considering the expected mechanical precision this would lead to a slightly worsened resolution of $11\mu$m.
In conclusion the study demonstrated that the simplest idea of an
half-strip staggering of the two sensor layers alone, to be realised in
hardware, would not provide a stable working point, while the
effect of a substantial tilt of the detector geometry is robust.

\begin{figure*}[!htbp]
\begin{center}
\includegraphics[scale=0.62]{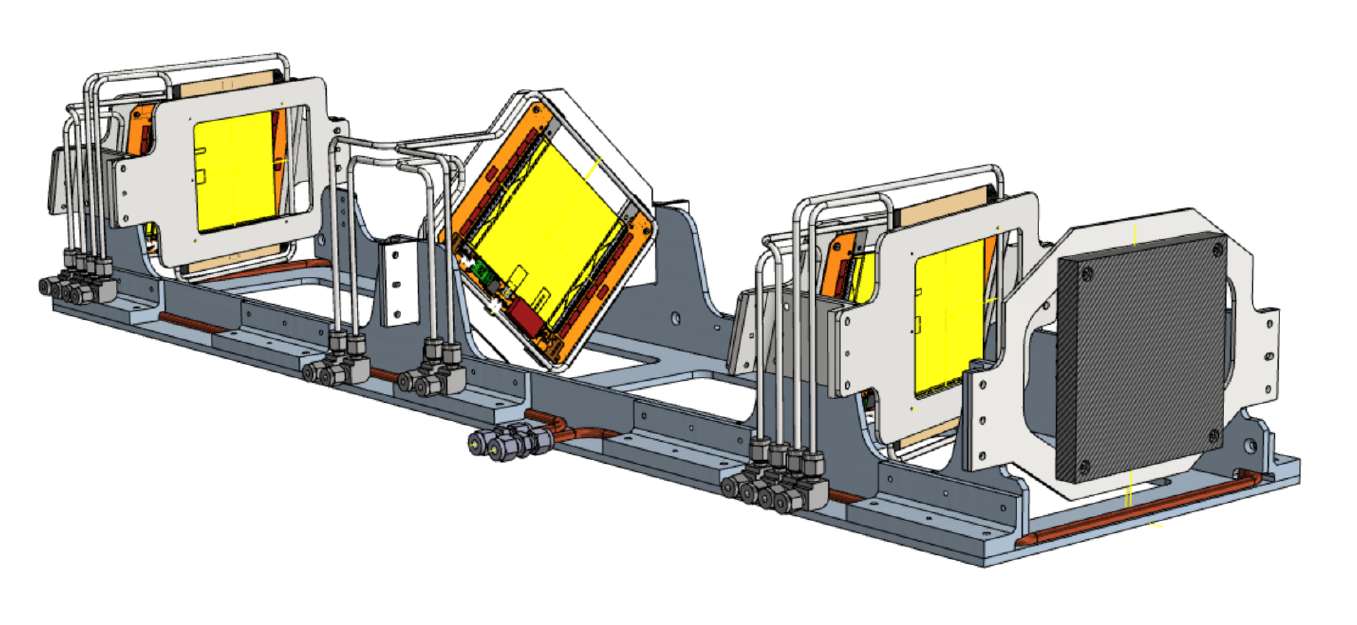}
\caption{CAD drawing of a MUonE tracking station.}
\label{MUonE-station-CAD}
\end{center}
\end{figure*}
As a consequence, the mechanical design of a MUonE tracking station has
been updated as shown in Figure~\ref{MUonE-station-CAD}.
The structure length is still one meter. 
The target plate is followed by three equally-spaced XY supermodules,
each one made of two close-by 2S modules with orthogonal strips. 
The first and last supermodules, measuring X and Y transverse coordinates,
have tilted modules by 233~mrad, as determined by the
study. The middle supermodule is rotated by 45$^o$ around the beam axis,
to resolve reconstruction ambiguities, and its modules are not tilted. 
There are stringent requirements on the mechanical stability of the
tracking stations, which has to be better than $10 \mu$m, in
particular on the longitudinal size.
Therefore the support structure is made of Invar (Fe-Ni alloy), which has a very low coefficient of thermal expansion, is
easy to machine and relatively cheap. 
A cooling system is also designed, in addition to an enclosure to
stabilise the room temperature within $1^o$-$2^o$.

\section{MUonE analysis technique}\label{sec:analysis}
The hadronic contribution to the running of $\alpha$ is most easily
displayed by considering the ratio $R_{\rm had}$ of a cross section
including the full running of $\alpha$ and the same cross section with no
hadronic running.
At Leading Order (LO) the cross section has simply the squared
coupling factorised, so this approximation holds:
\begin{equation}
R_{\rm had}^{LO}(t) = \frac{d\sigma^{LO}(\Delta\alpha_{\rm had}(t) \neq 0)}{d\sigma^{LO}(\Delta\alpha_{\rm had}(t) = 0)} \simeq 1 + 2 \Delta\alpha_{\rm had}(t).
\end{equation}
For a muon beam energy of 160~GeV, the centre-of-mass energy is
$\sqrt{s} = 0.418$~GeV and the maximum momentum transfer is $t_{max} =
-0.175$~GeV$^2$.
The hadronic contribution $\Delta\alpha_{\rm had}(t)$ has a tiny variation across the probed
kinematical range $0 < |t| < |t_{max}|$, changing from a
vanishingly small value at low $|t|$ to about $10^{-3}$ at the peak
of the integrand in (\ref{amu_xalpha}), which occurs at $t = -0.108
\mathrm{~GeV}^2$ ($x = 0.914$).
A competitive determination of \amuHLO~ requires a
precision of ${\cal O}(10^{-2})$ in the measurement of the hadronic
running, which translates into an unprecedented precision of ${\cal
  O}(10^{-5})$ in the shape of the differential cross section.
Reaching this accuracy requires a huge statistics of data, in the order
of few times $10^{12}$ events. Therefore even preliminary
simulation studies would present a computational challenge.
A smart trick applicable to the simulation studies
consists in using the very same MC sample to determine the
$R_{\rm had}$ ratios, by reweighting events to include or exclude the
hadronic component of $\alpha(t)$. The distribution simulating real
data (the \emph{pseudodata}) is determined with the hadronic running switched
on, and is fluctuated for the expected statistical uncertainties
corresponding to the desired luminosity.
The distribution representing the theory prediction is determined with the hadronic running switched
off and without additional fluctuations.
In this way the correlated uncertainties due to the limited number
of generated events will cancel out in the ratio of the two distributions, given that they differ just for
the slightly different event weights, and the signal from the hadronic running will
be visible.
Obviously this trick works only with the MC simulation,
while with real data one will have to match the generated MC
statistics to the number of real data. 

At Next-to-Leading Order (NLO) the ratios $R_{\rm had}$ can be defined for
observables like the scattering angles by using a more complex
reweighting technique, considering the structure of the matrix
element in events with the emission of a real photon. 
No attempt is done to estimate the momentum transfer $t$ event by event.
Actually radiative events strongly
modify the LO kinematics and their correct description is necessary. 
The analysis described in \cite{LoI}
used a NLO Monte Carlo generator implementing an exact calculation including
masses ($m_\mu, m_e$) and electroweak corrections in a fully
differential code \cite{NLOgen}.
The hadronic contribution
$\Delta\alpha_{\rm had}(t)$ is included in the generator by the Jegerlehner
numerical parameterisation \cite{Jeger95, Jeger08},
which is obtained from the dispersive integral of low-energy measurements of
hadronic production in $e^+e^-$ annihilation and perturbative QCD.
This parameterisation gives $a_\mu^{\rm{HVP,\,LO}} = 688.6 \times 10^{-10}$ when it is
inserted into the master integral (\ref{amu_xalpha}).
Other existing parameterisations can be used as well.

Beam spread and detector resolution effects are included in a fast
simulation. The muon beam is given a spread of $3.75\%$ around its nominal
value of 150~GeV with a Gaussian distribution.
The intrinsic angular resolution for the measured tracks is assumed to be
$\sigma_\theta = 0.02$~mrad.
The multiple Coulomb scattering is parameterised by a Gaussian with
the usual approximation for the width \cite{MCS}.

The extraction of the hadronic contribution is carried out by a
template fit method, in which the templates for the
observed distribution are calculated by reweighting the MC events to correspond
to an appropriate functional form of $\Delta\alpha_{\rm had}(t)$.
Several analytical forms have been tested.
A third order polynomial could in principle describe
the MUonE data but is unphysical, and it would make the integrand in
(\ref{amu_xalpha}) divergent for $x \to 1$.
A Pad\'e approximant with three free parameters, like:
\begin{equation}
\Delta\alpha_{had}(t) = a t \frac{1 + bt}{1 + ct}
\label{pade}
\end{equation}
would be well behaved.
However the best found option is a
physically inspired parameterisation, corresponding to the one-loop QED
calculation of vacuum polarisation induced by a lepton pair in the
space-like region:
\begin{equation}
\fl
\Delta\alpha_{had}(t) = k \left\{ -\frac{5}{9} -\frac{4M}{3t} +
\left( \frac{4M^2}{3t^2} + \frac{M}{3t} -\frac{1}{6} \right)
\frac{2}{\sqrt{1-\frac{4M}{t}}} \mathrm{log} \left|
\frac{1-\sqrt{1-\frac{4M}{t}}}{1+\sqrt{1-\frac{4M}{t}}} \right|  \right\},
\label{DalphaLLparam}
\end{equation}
where the $M$ parameter replaces the squared lepton mass $m^2$ and $k$
the factor $\alpha/\pi$. The same form is also valid for the contribution of $t\bar{t}$
pairs (with $M = m_{top}^2$ and $k = \frac{\alpha}{\pi} Q^2 N_c$, with
$Q=2/3$ the top electric charge and $N_c = 3$ the number of colours).
Since the hadronic contribution to the running $\alpha$ is
not calculable in perturbation theory, the parameters $k$ and $M$
do not have a precise physics interpretation. 
At large $|t|$ the dependency is logarithmic, proportional to
$\mathrm{log}(|t|/M)$ as expected.
In the limit of very small $t$ it reduces to a linear trend:
\begin{equation}
\Delta\alpha_{had}(t) \simeq - \frac{1}{15} \frac{k}{M} t,
\label{DalphaLLlowt}
\end{equation}
which corresponds to the dominant behaviour in the MUonE kinematical region. Correcting
terms, corresponding to quadratic and higher orders in $t$ are
incorporated in the form of (\ref{DalphaLLparam}).  

The \emph{Lepton-Like} parameterisation (\ref{DalphaLLparam}) has been
preliminarily tested against the Jegerlehner numerical parameterisation.
The level of agreement is excellent, considering that (\ref{DalphaLLparam}) has only two free
parameters. In comparison the Pad\'e form (\ref{pade}) cannot fit equally well the Jegerlehner
parameterisation. In addition, having three instead of two free
parameters leads to larger uncertainties in the resulting integral
(\ref{amu_xalpha}).

\begin{figure*}[!htbp]
\begin{center}
\includegraphics[scale=0.4]{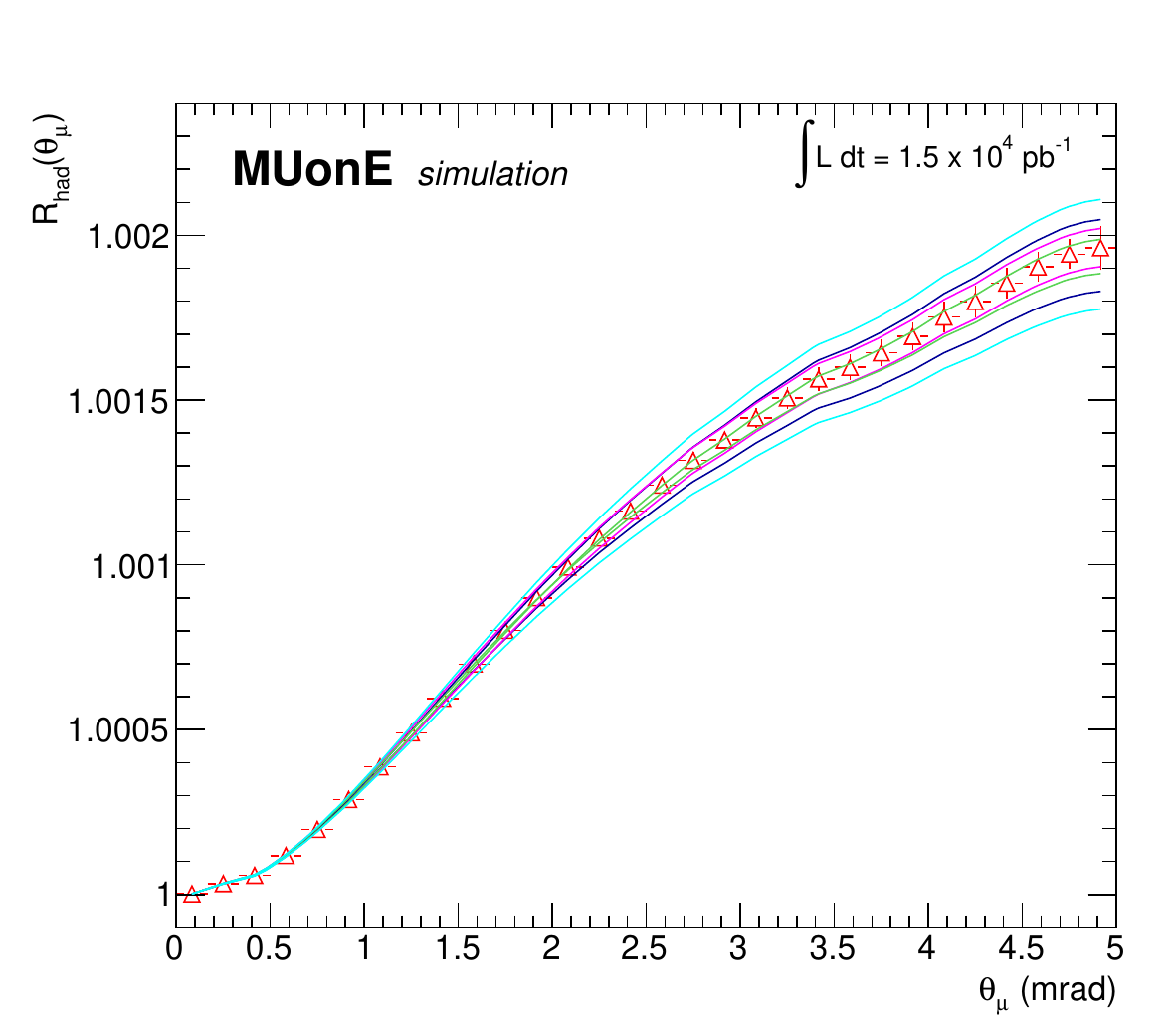}
\caption{Central prediction for $R_{\rm had}(\theta_\mu)$. The error bars
  correspond to the expected statistical uncertainties for the nominal MUonE
  luminosity of $1.5 \times 10^4$~pb$^{-1}$. The curves show a few representative MC templates.}
\label{templates}
\end{center}
\end{figure*}

The template fit is carried out by defining a grid of points $(k,M)$
in the parameter space covering a region of $\pm5\sigma$ around the
expected values, with $\sigma$ being the expected uncertainty. 
Actually, due to the dominant low-$t$ dependence, the $k$ and $M$
parameters are highly correlated and it is convenient to use $K=k/M$ as fit
parameter, substituting $k = KM$ in (\ref{DalphaLLparam}).
The step size is taken to be $0.5\sigma$. This defines $21 \times 21 =
441$ templates for the relevant distributions. Figure~\ref{templates}
shows a few representative templates in the chosen window for the
angular distribution of the scattered muon, together with the central
values expected for the MUonE nominal luminosity.
Every template in the grid is compared to the pseudodata calculating:
\begin{equation}
\chi^2(K,M) = \sum_{i} \frac{R_i^{data} - R_i^{templ}(K,M)}{\sigma_i^{data}},
\end{equation}
where the sum runs over all the bins,
and the minimum $\chi^2$ is found by parabolic interpolation across
the grid points. The error is determined for $\Delta\chi^2 = 1$.

The fit can be done on the distribution of the muon or the electron
scattering angle, as well as on their two-dimensional distribution,
which gives the most accurate result.
Actually there is almost no need to identify the outgoing muon and
electron, provided the event is a signal one. In this case the two
track angles are labelled as $\theta_L$ and $\theta_R$, meaning \emph
{Left} or \emph{Right} with respect to an arbitrary axis.

\begin{figure*}[!htbp]
\begin{center}
\includegraphics[scale=0.38]{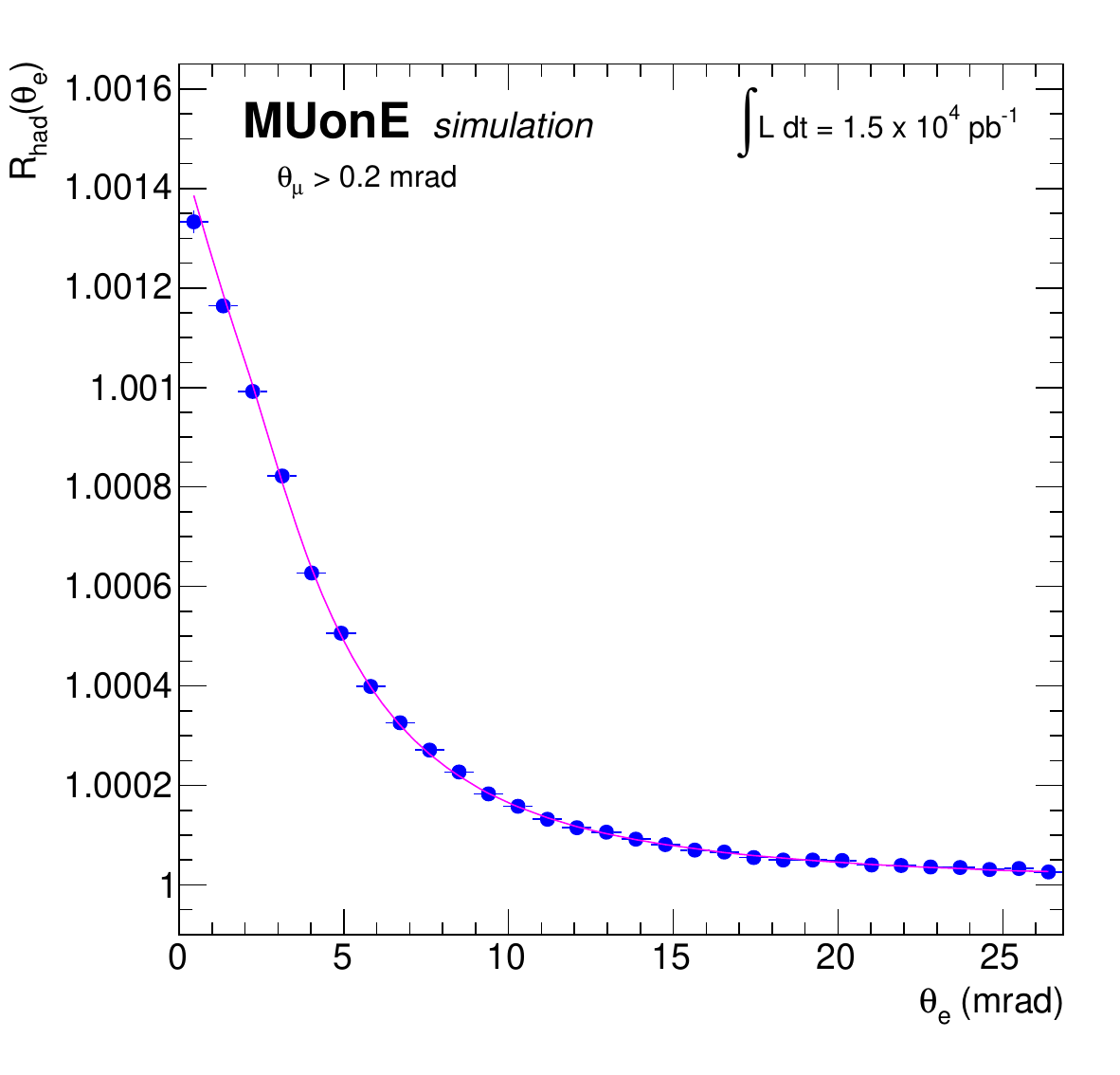}
\includegraphics[scale=0.38]{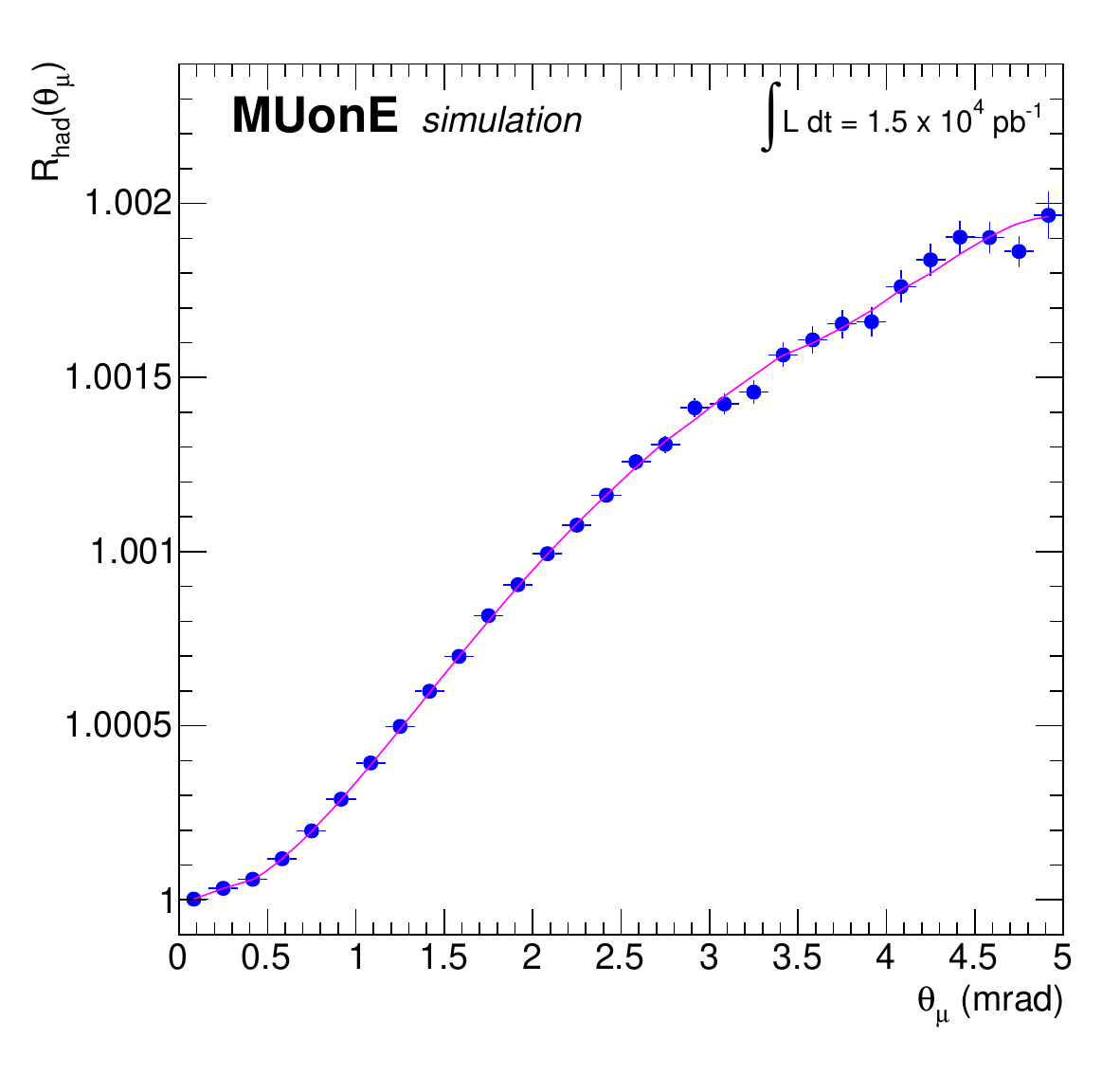}
\caption{Example pseudodata showing the ratios $R_{\rm had}$ for
  (\emph{Left}) the electron angular distribution, after a cut at
  $\theta_\mu > 0.2$~mrad, and (\emph{Right}) the muon angular
  distribution. The error bars show the statistical
  uncertainties corresponding to the nominal MUonE luminosity. The template
  fit is superimposed.}
\label{fits_1dproj}
\end{center}
\end{figure*}

The geometric acceptance of the MUonE stations covers the
relevant region of scattering angle $\theta \leq 32$~mrad, which in LO
corresponds to outgoing electrons with energy greater than
1~GeV. 
It is important to remind that only the shape of the angular
distributions is relevant, the
absolute normalisation is neglected, as it will suffer from a relatively
large systematic uncertainty related to the luminosity determination.
The fitted parameterisation (\ref{DalphaLLparam}) is then inserted into the
master integral (\ref{amu_xalpha}), and the value of \amuHLO ~is
determined by integrating over the full phase space.
Figure~\ref{fits_1dproj} shows the $R_{\rm had}$ distributions
obtained for the muon and electron angle,
for an example pseudoexperiment, with the fit result superimposed.
The statistical accuracy of the fit has been tested by repeating many
pseudoexperiments, each one with statistics equivalent to the MUonE
nominal luminosity. From 1,000 pseudoexperiments we get 
$a_\mu^{\rm{HVP,\,LO}} = 688.8 \pm 2.4 \times 10^{-10}$ which is in very good
agreement with the expected value of the Jegerlehner parameterisation
used in the generator ($688.6 \times 10^{-10}$). With respect to the result
obtained in \cite{LoI} we have improved the fit technique, basically
removing the systematic uncertainty related to the fit model which we
had mentioned there. Further studies will be carried out using other
input models\footnote{An independent procedure to determine $a_\mu^{\rm{HVP}}$ from
  MUonE measurements alone, integrating over the full phase space, has
  been published very recently \cite{Greynat-deRafael}.}.

\section{Test Run}\label{sec:TestRun}
A first important milestone is a three-week Test Run
at the CERN M2 beam line, with full intensity muon
beam. The selected location is upstream of the COMPASS detector,
after its Beam Momentum Spectrometer (BMS) \cite{COMPASS:2007}.
The MUonE setup will consist of two tracking stations followed by the ECAL,
with an additional station (without target) upstream to track the
incoming muons \cite{LoI}.
Initially the Test Run was allocated in fall 2021.
However there were some delays in the procurement of the
module components and a shift in the CMS preproduction timeline,
so the final schedule has to be redefined.

This forthcoming Test Run shall confirm the detector design and
engineering and constitute a proof of concept for the many challenges
to be faced. Among these, the procedures for the alignment (hardware
and software), the readout chain and the trigger strategy to identify
and reconstruct $\mu e$ events.

The physics potential of the Test Run setup has been estimated by
considering the standard SPS efficiency with full beam intensity, and
allowing for the necessary time for detector commissioning \cite{ICHEP20}.
The two stations could yield $\sim$1~pb$^{-1}/$day, and reasonably we
could integrate up to $\sim$5~pb$^{-1}$ of good data
during a first physics run, with a safety margin for possible data-taking inefficiencies.
This integrated luminosity would correspond to $\sim$10$^9$ $\mu e$
scattering events with electron energy greater than 1~GeV.

\begin{figure*}[!hbtp]
\begin{center}
\includegraphics[width=.5\textwidth]{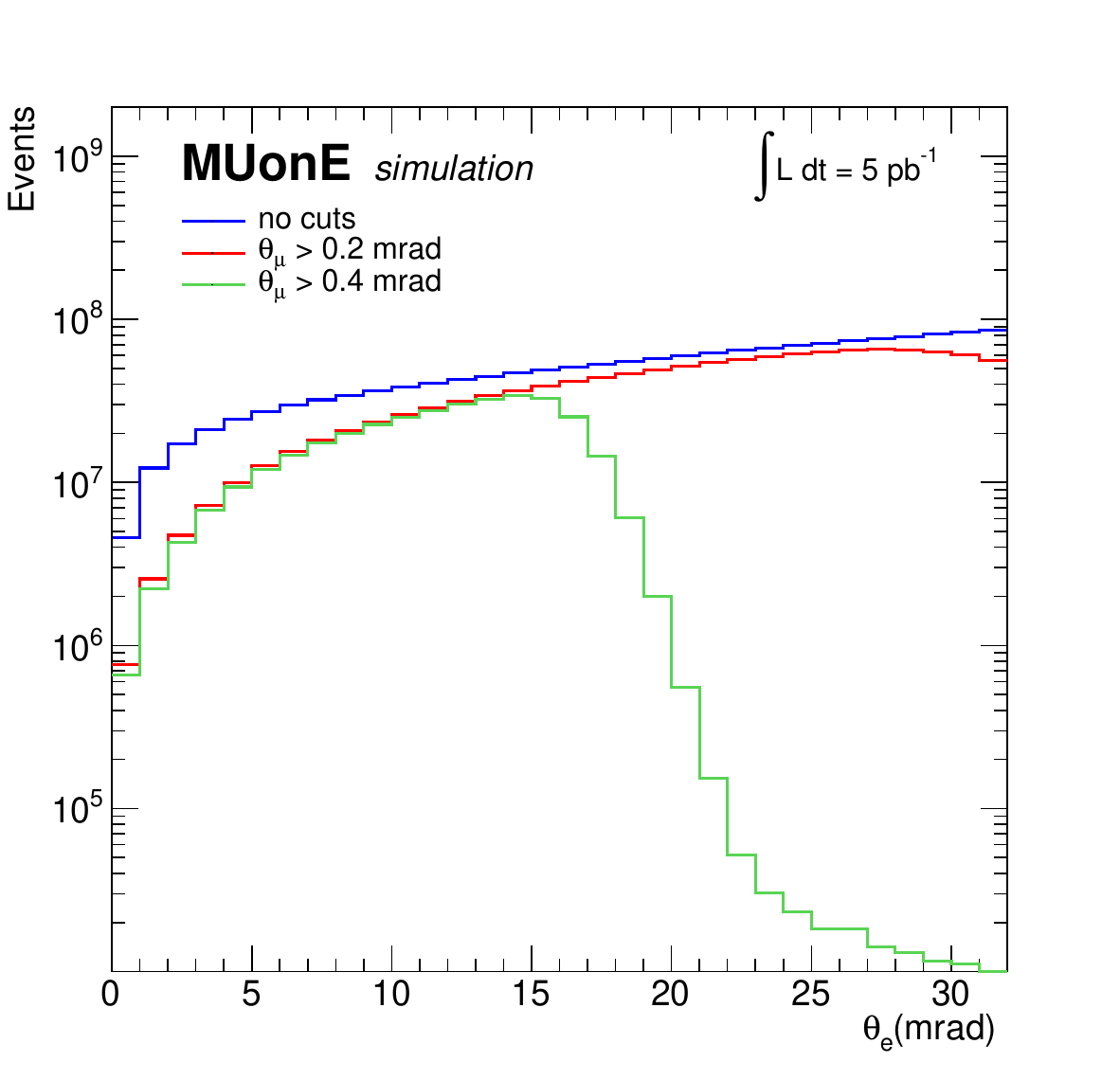}~\includegraphics[width=.5\textwidth]{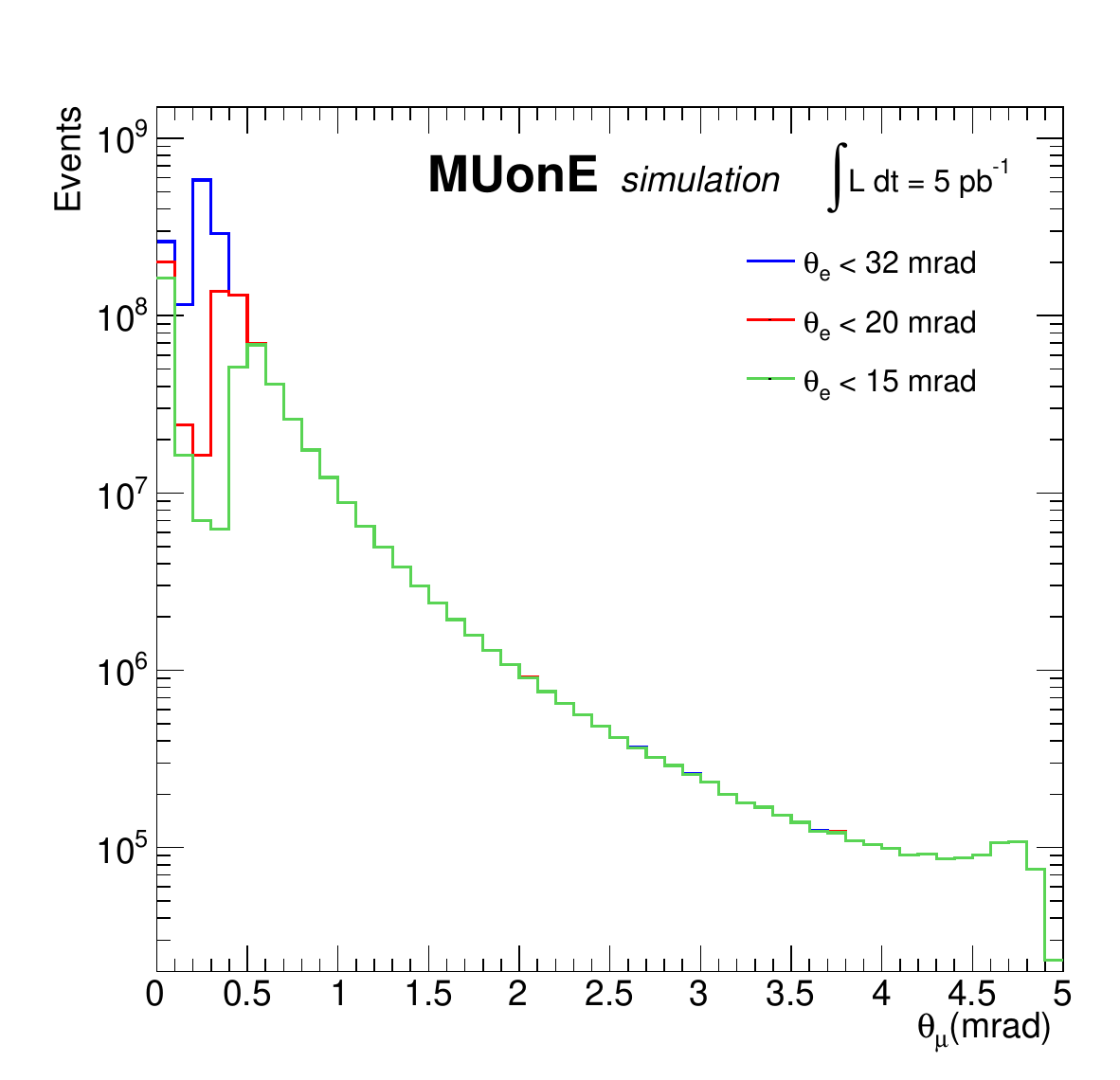}
\caption{Event yields expected in the Test Run, for: (\emph{Left}) the
  electron and (\emph{Right}) the outgoing muon, after the application
  of simple angular cuts.}
\label{nevents_TR}
\end{center}
\end{figure*}

The expected event yields are shown in Figure~\ref{nevents_TR} for the
electron and muon angular distributions.
The muon distribution has a dip at very
small angles, which results from the $\theta_e < 32$~mrad geometric
acceptance for the outgoing electron. Tighter cuts on the electron angle produce a wider dip on
the muon distribution. Muons entering the dip region result from
events with one real photon emitted, which smear the perfect correlation expected for
elastic events.
A minimal cut $\theta_{\mu} > 0.2$~mrad, aimed to reject radiative events,
strongly affects the shape of the electron angular distribution, as shown
in Figure~\ref{nevents_TR}~(\emph{Left}).
The removed events are characterised by low-energy electrons which would contaminate the signal
region at small $\theta_e$. By applying a tighter cut on the muon
angle, as $\theta_{\mu} > 0.4$~mrad, the electron distribution is cut
for angles greater than $15-20$~mrad. The visible edge is smeared by detector
resolution and radiative events.

\begin{figure*}[!hbtp]
\begin{center}
\includegraphics[width=.5\textwidth]{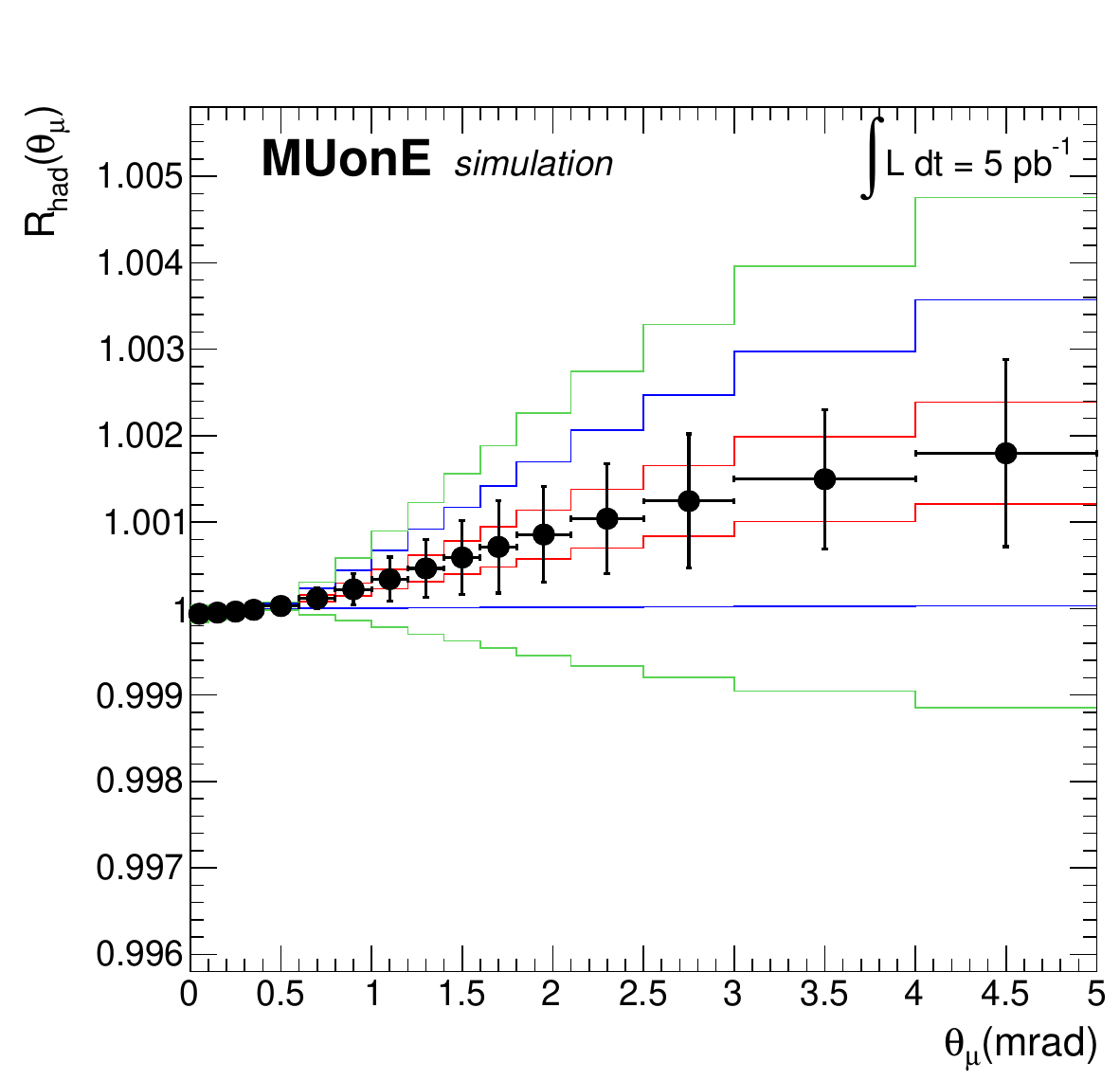}
\caption{Ratio $R_{\rm{had}}$ of the expected muon angular
  distribution and the prediction obtained with only leptonic
  running in $\alpha(t)$. The error bars correspond to the statistical
  uncertainties for an integrated luminosity of 5~pb$^{-1}$ assumed
  for the Test Run. The histograms show the templates for few values of the slope
  $K$. Plot from \cite{ICHEP20}.
}
\label{TR21_fits}
\end{center}
\end{figure*}

The statistics achievable at the Test Run would give enough sensitivity to measure the leptonic running
of $\alpha$ and potentially could provide initial sensitivity to the hadronic running.
With respect to the nominal luminosity, the fit method has been
adapted to a simpler case.
The parameter $M$ in (\ref{DalphaLLparam}) is fixed to its
expected value $M=0.0525$~GeV$^2$ and only $K = k/M$ is fitted, to determine a
linear deviation on the shape. Fig.~\ref{TR21_fits} shows the
expectation for the muon $R_{\rm had}$ distribution with a few template
distributions superimposed. 
From 1,000 pseudoexperiments the central value of the fit is found to
be $K = 0.136 \pm 0.026$.
This represents the slope of the observable
hadronic running at the purely statistical level. 

It is important to assess the expected systematic errors, which will
have to be fully understood in the Test Run conditions to estimate
their impact in the full-scale experiment.
The most important systematic effects will be measured from the data
itself, in the so-called \emph{normalisation region}, where the hadronic
running is negligible. This corresponds to events where the muon is scattered at small
angle keeping most of its initial energy, while the emitted electron
goes at relatively large angles with energy of few GeVs. The recorded statistics in this
region will be very large allowing for very precise measurements of the
detector performance.

\begin{figure*}[!hbtp]
\begin{center}
\includegraphics[width=.5\textwidth]{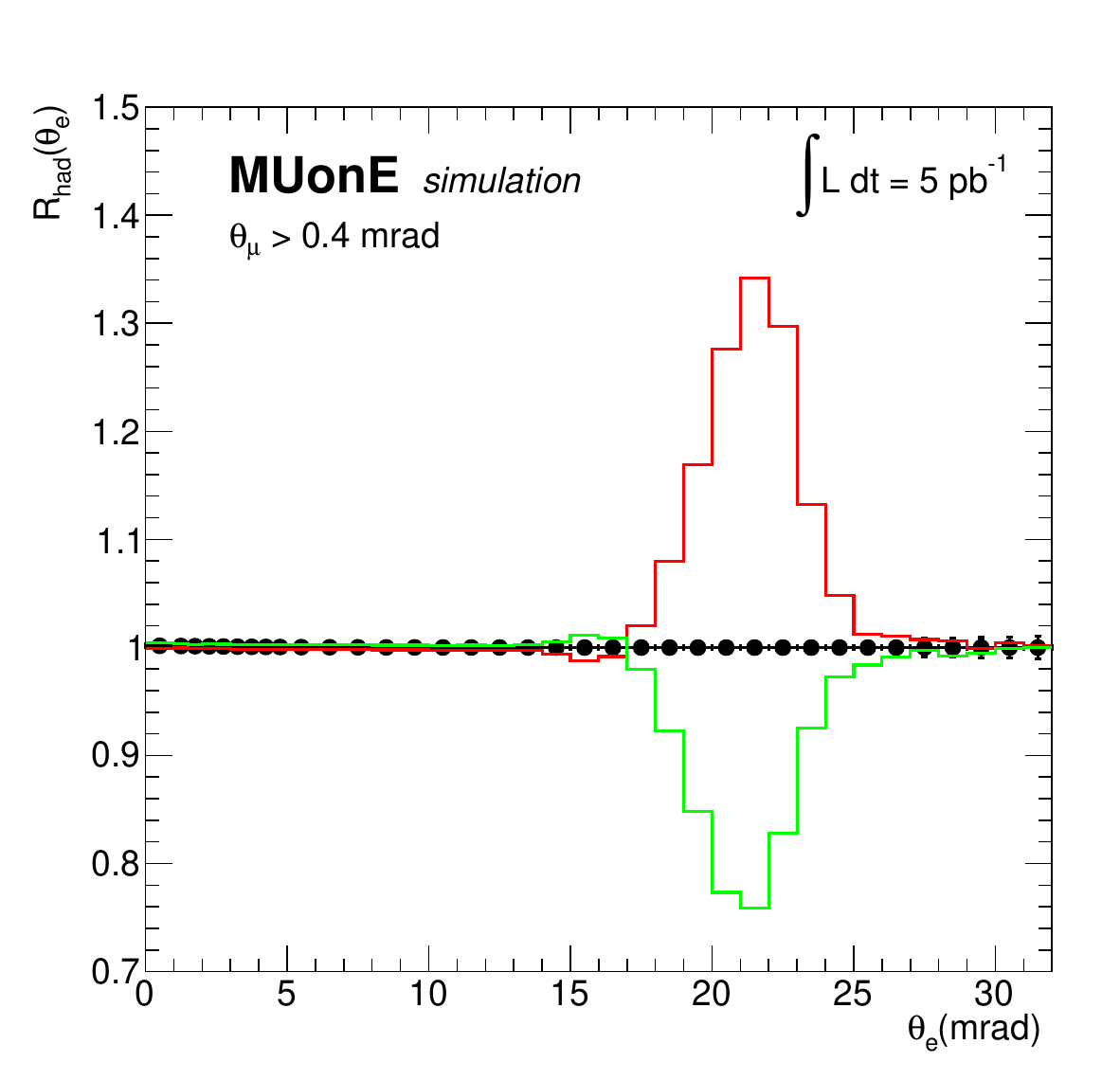}~\includegraphics[width=.5\textwidth]{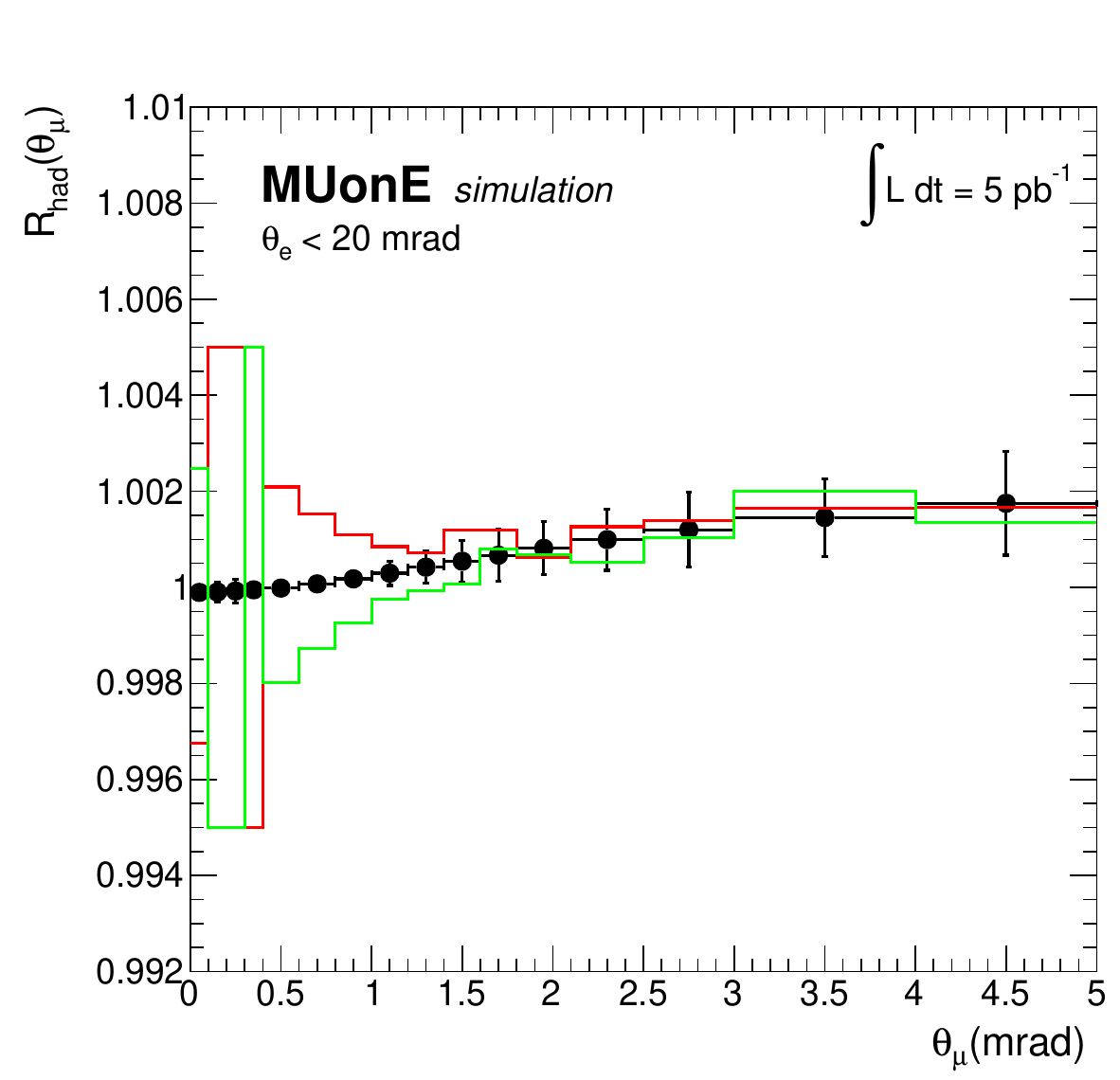}
\caption{Effect of a systematic error of $+10\%$~(red
    histograms) or $-10\%$~(green histograms) in the intrinsic angular
  resolution of the detector (expected $\sigma_\theta$~=~0.02~mrad), as visible in the $R_{\rm had}$ ratios:
  (\emph{Left}) the electron distribution after a cut at 
$\theta_{\mu} > 0.4$~mrad; (\emph{Right}) the muon distribution, after
a cut at $\theta_e < 20$~mrad. The solid points represent the expected
central values with error bars showing the statistical uncertainties
for the Test Run.}
\label{syst_intr_resol}
\end{center}
\end{figure*}

The intrinsic resolution
is one of the most important figures to be determined. It will be
routinely monitored while carrying out the track-based detector
alignment. However a more precise measurement is required by
the physics analysis. 
A sharp control over this parameter is possible by applying a cut on the
muon angle at $\theta_{\mu} > 0.4$~mrad and observing the shape of the
resulting edge in the electron angle distribution. Figure~\ref{syst_intr_resol}~(\emph{Left})
shows the expected distortion which would be visible for a $\pm10\%$
systematic error on the intrinsic angular resolution.
Reversing the roles, Figure~\ref{syst_intr_resol}~(\emph{Right}) shows
the effect which would be seen in the muon angle distribution after
applying a cut on the electron angle at  $\theta_{e} < 20$~mrad.

\begin{figure*}[!hbtp]
\begin{center}
\includegraphics[width=.5\textwidth]{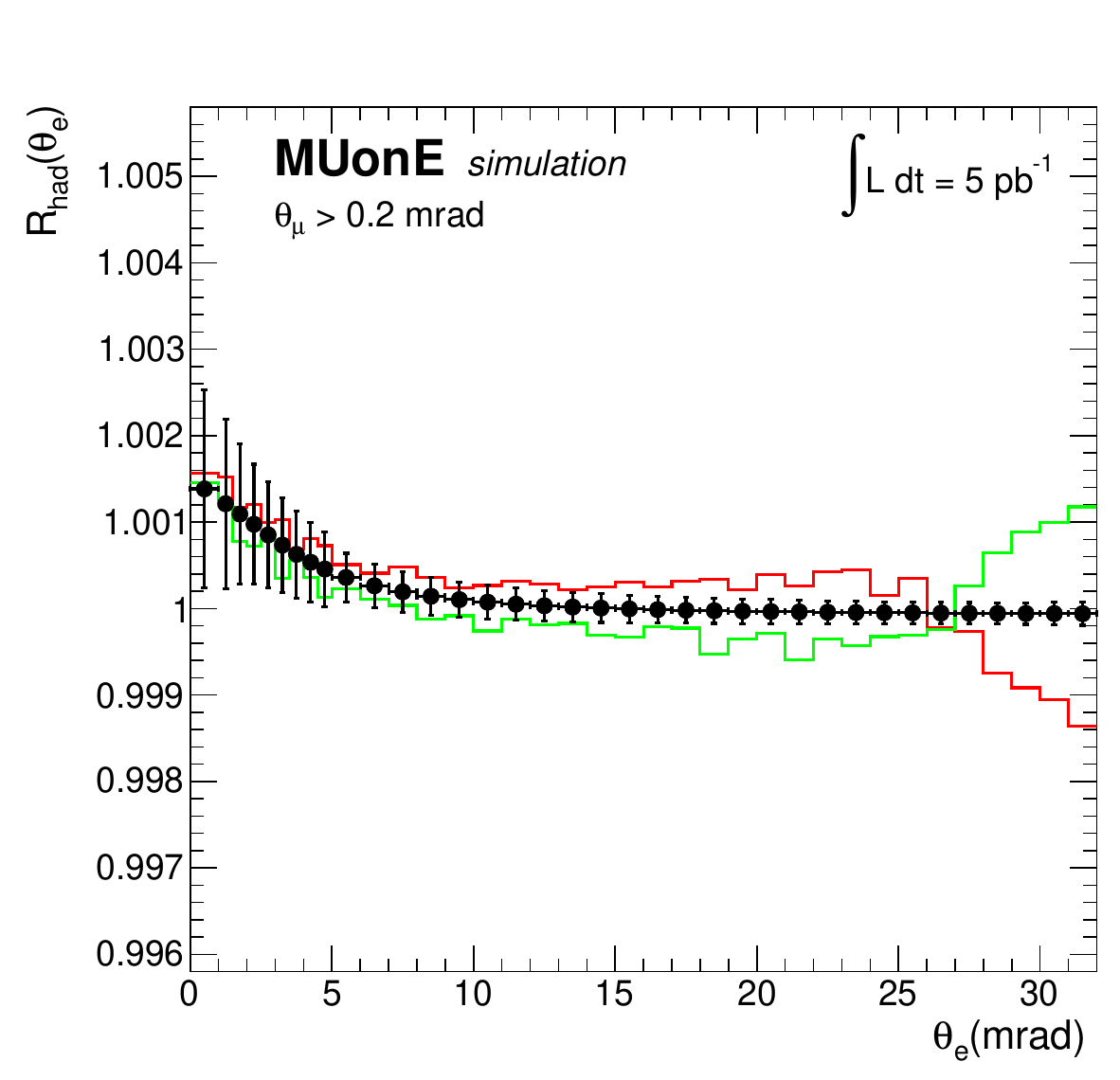}~\includegraphics[width=.5\textwidth]{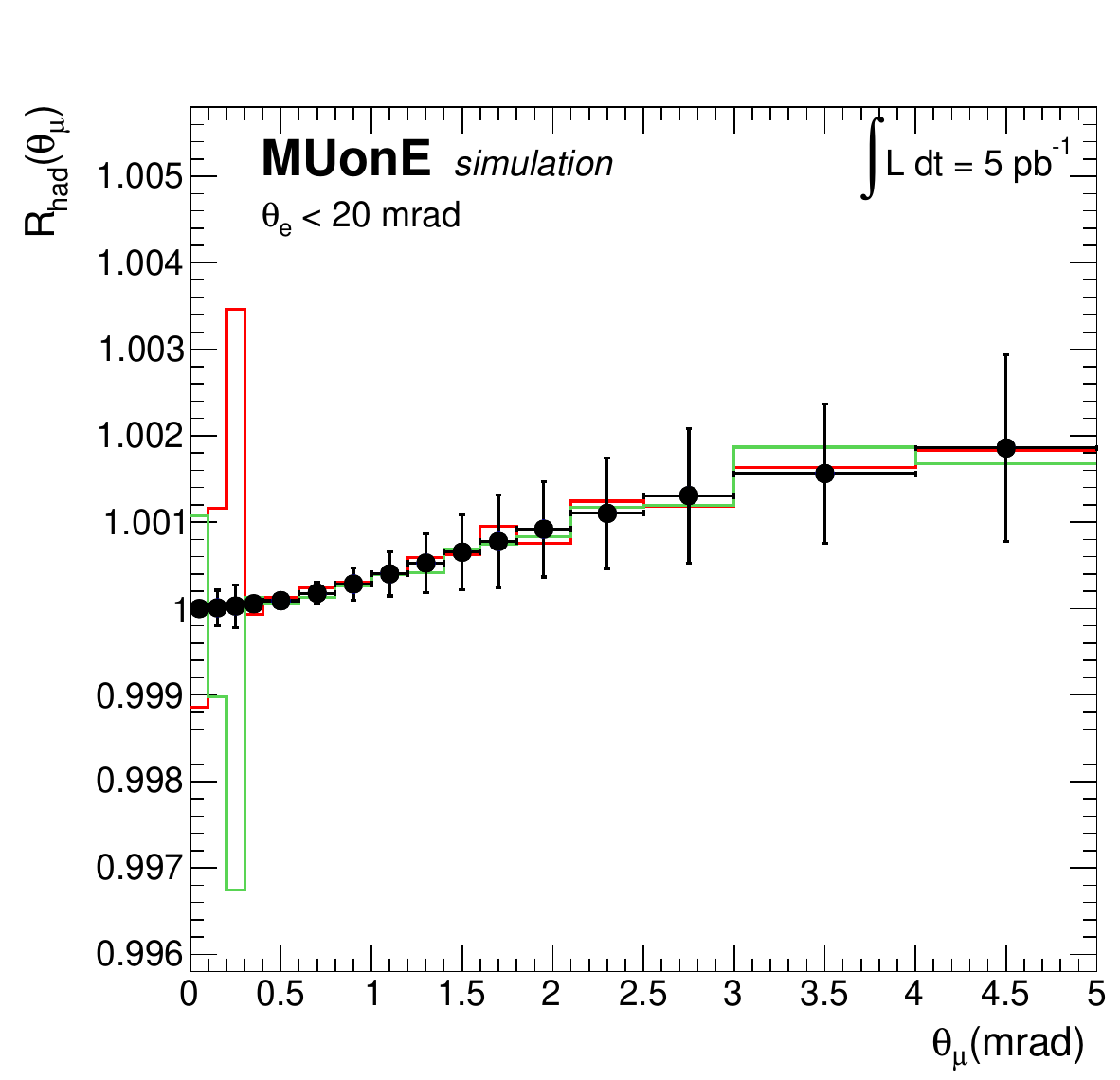}
\caption{
Effect of a systematic error of $+1\%$~(red histograms) or $-1\%$~(green histograms) on the assumed
  width of the core of the MCS distribution, as seen in the
  $R_{\rm{had}}$ ratios: (\emph{Left}) the electron distribution
  after a cut at $\theta_{\mu} > 0.2$~mrad,
  and (\emph{Right}) the muon distribution (from \cite{ICHEP20}). 
The solid points represent the expected
central values with error bars showing the statistical uncertainties
for the Test Run.
}
\label{syst_MS}
\end{center}
\end{figure*}

The final detector resolution will depend also on the material
effects, in particular the multiple Coulomb scattering (MCS), mostly affecting the low energy
electrons. MCS of $12$~GeV and $20$~GeV electrons on thin carbon targets
has been studied in a dedicated beam test in 2017 \cite{TB2017}.
The core of the measured angular distributions was found to agree within $\pm
1\%$ with the predictions from the GEANT4 simulation.
The effect of a flat $\pm1\%$ error on the MCS core width, respectively on the muon and the electron angular
distributions is shown in Figure~\ref{syst_MS}.

The observable patterns produced by systematic errors on the detector
resolution are so evident 
that a first natural step will consist in a calibration of these
effects by fitting the data in control regions.
After this first calibration the residual systematic uncertainties will be included as
nuisance parameters in a likelihood fit and will be determined
simultaneously with the signal parameters.
We have tested the capability of doing such a simultaneous fit by using the
{\tt combine} tool \cite{combine, combine-web}, a software package used for
statistical analysis within CMS, based on RooStats/RooFit
\cite{roostats-roofit}. The two systematic effects related to the
detector resolution described above have been included, together with
a normalisation nuisance representing the luminosity uncertainty.
Angular distributions corresponding to the effects of $\pm 1 \sigma$ 
for each systematic source taken alone have been provided as input to the tool,
for every value of the physics parameter $K$. Then, for each of these
$K$ values the nuisance parameters are fitted from the pseudodata. 
The best fit for the signal parameter $K$ is then
found by parabolic interpolation over the grid points. 
Since the nuisance parameters are weakly correlated with the physics
parameter $K$, their final values could also be easily approximated by
interpolation. Otherwise, if needed, one could do a second step, by fixing the $K$
value to the best fit for it and producing $\pm 1 \sigma$ templates
around it to do this last minimisation.
The fit has proved to be reliable within the covered range
of uncertainties ($\pm 10\%$ for the intrinsic angular resolution and
$\pm 1\%$ for the core width of MCS). Input pseudodata with systematic
errors within this range have been successfully fitted, with almost no
degradation for the fitted signal parameter $K$.
As said, this works so well as the signal and the nuisances mostly act
on different kinematical regions.

\begin{figure*}[!htbp]
\begin{center}
\includegraphics[scale=0.4]{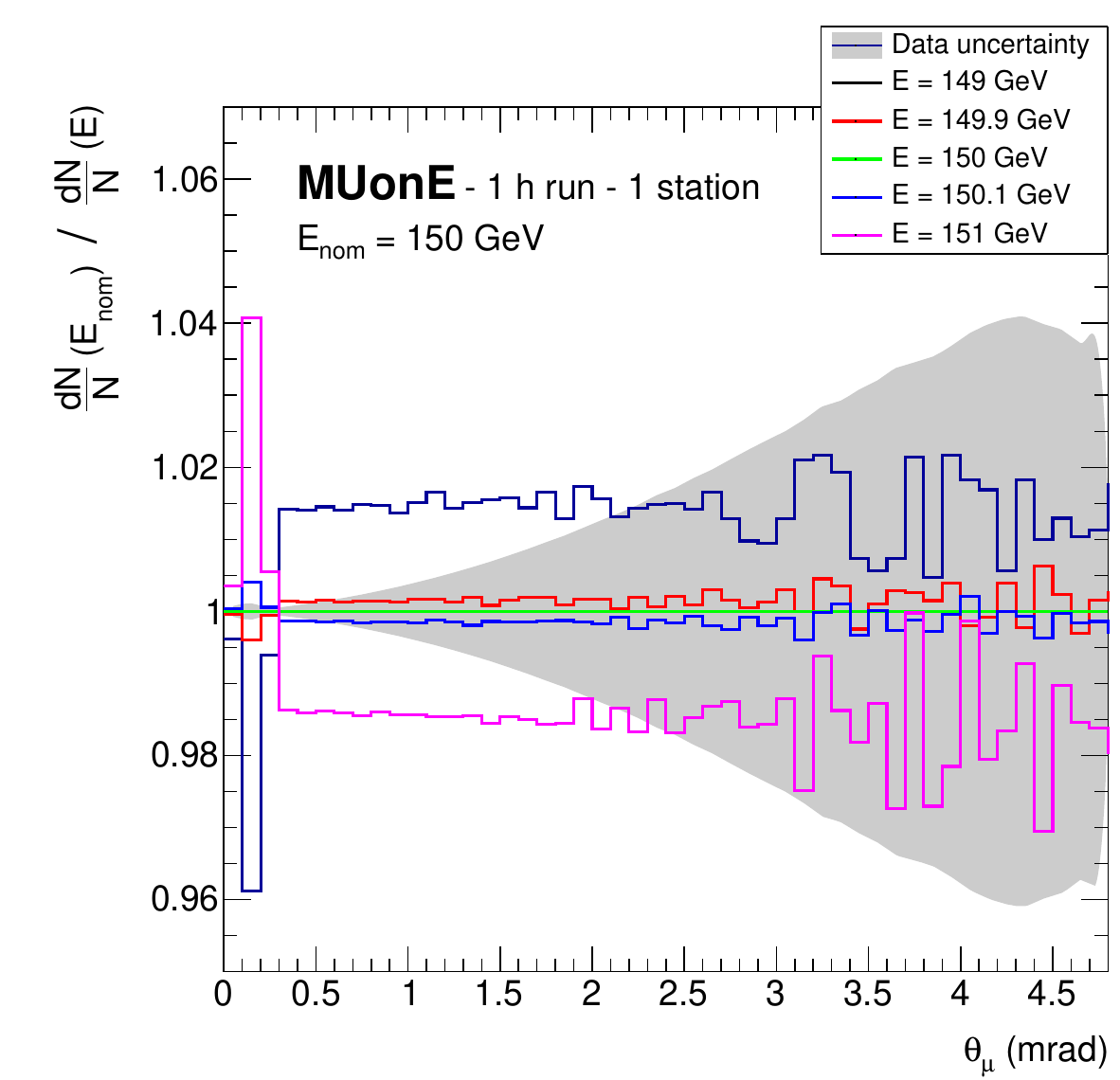}
\caption{Effect of a shift of $\pm (0.1$-$1.0)$~GeV in the average
  beam energy with respect to the nominal value $E_{\rm{nom}}=150$~GeV. The
  histograms show the expected distortions on the muon angular
  distribution, obtained from MC samples with variable energy. The grey band represents the statistical uncertainty
  corresponding to the expected data collected in one hour running
  time by one station. Plot from \cite{ICHEP20}.
}
\label{TR21_sys_escale}
\end{center}
\end{figure*}

Another crucial systematic effect is related to the knowledge of the
average beam energy scale. This is known from the accelerator at a
level of about 1\%. The BMS spectrometer can measure individual incoming muons with
0.8\% resolution, and given the high muon beam intensity it can
provide an excellent monitor of time variations of the average scale.
However it cannot assess the systematic uncertainty of the average
energy scale, which has to be controlled by a physical process.
The kinematics of the elastic $\mu e$ scattering has been identified
as the useful method \cite{LoI}, in particular the average angle of
the two outgoing tracks, which does not need $\mu$-$e$ identification.
For illustration purpose Figure~\ref{TR21_sys_escale} shows the effect
on the muon angular distribution
of a systematic error of $\pm (0.1$-$1.0)$~GeV on the
assumed average beam energy.
The expected distortion is compared to the statistical uncertainty corresponding to
one hour running time in one station.
It is clear that the energy calibration by the kinematical method would already outperform
the precision of the scale obtained from the accelerator.
The beam energy scale will be calibrated on each tracking station
independently, aiming at an ultimate precision for the final detector better than 3 MeV in less
than one week of run.

\section{Theoretical progress}\label{sec:theory}
An intense theoretical activity has been going on in the last years,
to obtain the very precise calculations needed for the MUonE measurement.
A comprehensive review \cite{muone-theory-initiative} describes
the full NLO calculation, including both QED and electroweak
corrections, also made available in a fully exclusive MC event
generator \cite{NLOgen},
the calculation of NNLO hadronic corrections \cite{Fael:2018dmz, Fael:2019nsf},
and the evaluation of the two-loop integrals relevant for the NNLO QED
corrections \cite{Mastrolia:2017pfy, DiVita:2018nnh, DiVita:2019lpl}.

Very recently, the analytic evaluation of the two-loop corrections
to the process $e^+e^- \to \mu^+ \mu^-$ has been completed
\cite{Bonciani:2021okt}, treating the electron (muon) as massless
(massive) particle.

Another very recent paper \cite{Balzani:2021del} presented simple
exact analytic expressions to compute the hadronic vacuum polarisation
contribution to the muon $g-2$ in the space-like region up to
next-to-leading order. These results can be employed by MUonE to
extend the determination of $a_{\mu}^{\rm HVP}$ from leading to next-to-leading
order.

Moreover, the exact NNLO
photonic corrections on the leptonic legs, including all mass terms,
have been implemented in two independent and fully exclusive MC codes
\cite{MCNNLO_PV,MCNNLO_PSI}.
Their results have been compared and are in very good agreement.
Resummation of leading terms in higher orders (by parton shower and
YFS exponentiation) matched to (N)NLO will be necessary and is
being carried out.

In addition, the complete NNLO virtual and real leptonic corrections have
been calculated and implemented in the MESMER MC code \cite{MC-pair-prod}.
The real contributions constitute background events with the production of an $e^+e^-$ pair. 
The extraordinary accuracy requested by the MUonE selection demands also precise estimates of
the expected backgrounds. Very recently the contribution due to the emission of a
neutral pion has been studied \cite{PVpi0}.

Finally, possible contaminations from new physics effects of the MUonE
measurement of $a_{\mu}^{\rm HVP}$ via elastic $\mu e$ scattering have
been studied and found to be below the detection sensitivity \cite{Masiero:2020vxk, Dev:2020drf}.
Despite this fact, it has been recently pointed out that complementary
selections of inelastic events at MUonE
could have strong sensitivity to light dark matter mediators, in the
mass range $10-200$~MeV, which could explain the muon $g-2$ anomaly.
In particular a dark $Z'$ predicted in $L_\mu- L_\tau$ gauge model
could be produced through the process $\mu e \to \mu e Z'$ \cite{asai21}.
Another promising channel would be the production of dark photon
through the process $\mu e \to \mu e A'$ followed by the
decay $A' \to e^+e^-$, which could be detected by reconstructing the
displaced vertex \cite{galon22}.

\section{Conclusions}\label{sec:conclusions}
The MUonE experiment could help understanding the puzzle of muon $g-2$,
by providing a third way to determine the leading order hadronic
contribution $a_{\mu}^{\rm HVP}$, independent of the traditional method using the dispersive
integral of time-like measurements and of the lattice QCD
calculations.
The MUonE method needs the measurement with unprecedented
precision of the shape of the differential cross section of $\mu e$ elastic scattering,
using the intense muon beam available at CERN, with energy of 160 GeV,
off atomic electrons of a light target.
The MUonE expectation of $0.35\%$ statistical uncertainty on
$a_{\mu}^{\rm HVP}$ constitutes a competitive benchmark, and demands a
strict control on all the systematic uncertainties to be reached. 

The availability of precise calculations will be
important in the next years, with the improving precision expected
from the Fermilab $g-2$ experiment and the J-PARC project.
We have briefly summarised the main theoretical achievements relevant to MUonE,
which resulted from steady developments in the last few years.
MUonE relies on the availability of state-of-the-art calculations
implemented in a fully exclusive Monte Carlo generator.
The needed target is a Monte Carlo code including the complete NNLO calculation
matched to the resummed contributions of leading logarithmic terms at all
orders of perturbation theory.
Moreover, precise predictions are needed also for the expected
background processes.
Further interest, in addition to the original MUonE motivation, has been
suggested by recent studies, indicating that in complementary phase space regions, excluded from the main selection,
MUonE could have competitive sensitivity to possible light dark matter
mediators, in the mass range $10 - 200$~MeV, which could explain the muon $g-2$ anomaly.

The experimental activities are progressing in preparation of the
approved Test Run with a reduced setup, which has to confirm the detector
design and verify the capability to reach the requested precision.
First estimates indicate that in few days
of data-taking in nominal conditions one could measure the leptonic
running of $\alpha$, with initial sensitivity to the hadronic running.
Tests with few detector modules have started in fall
2021. The foreseen setup will be integrated and tested as soon as the
hardware components become available, compatibly with the SPS schedule.

Meanwhile the detector design has been optimised with respect to the setup
described in the Letter of Intent. The new geometry of the
tracking stations can improve the hit spatial resolution by more than a factor of 2.

Also the analysis has been refined, removing a systematic
uncertainty related to the fitting technique. The template fit with
the two-parameter Lepton-Like parameterisation of the hadronic running
works very well, allowing for an unbiased extrapolation of the MUonE
measurement to the full integral giving $a_{\mu}^{\rm HVP}$.

Some of the most important expected systematics have been studied
by a fast detector simulation, obtaining encouraging results.
The detector resolution has been simulated by smearing the particle angles with a simple model with two
Gaussian components, the intrinsic angular resolution and the multiple Coulomb scattering.
The distortions resulting from errors in the assumed Gaussian widths can be fitted with high statistical accuracy
from the data itself, in control regions where the signal (the hadronic running of $\alpha$) is vanishing.
Another crucial systematic is related to the knowledge of the average beam energy
scale. A calibration method based on the event kinematics has been developed, 
that is expected to reach an ultimate precision of few MeVs for the final detector in less than one week of run.

A full experimental proposal will be prepared after the Test Run completion, assuming it to
be successful. The full detector construction could then take place, with the
prospect of a substantial running time during the LHC Run3, before the
start of the Long Shutdown scheduled in 2026-2028.

\end{document}